\documentclass[prb,twocolumn,aps,showpacs, superscriptaddress]{revtex4}

\usepackage{graphicx}
\usepackage{bm}
\usepackage{amssymb}
\usepackage{amsmath}
\usepackage{subfigure}

\newcommand{\kb}{{\boldsymbol{k}}}
\newcommand{\rb}{{\boldsymbol{r}}}

\begin{document}

\title{Variational Monte-Carlo studies of Gossamer Superconductivity}
\author{Siegfried Guertler}
\affiliation{Center of Theoretical and Computational Physics and Department of
Physics, The University of Hong Kong, Hong Kong, China\\}
\author{Qiang-Hua Wang}
\affiliation{National Laboratory of Solid State Microstructures \&
Department of Physics, Nanjing University, Nanjing 210093,
China\\}
\author{Fu-Chun Zhang}
\affiliation{Center of Theoretical and Computational Physics and Department of
Physics, The University of Hong Kong, Hong Kong, China\\}
\date{\today }

\begin{abstract}
We use a partially Gutzwiller projected BCS d-wave wavefunction
with an antiferromagentic weighting factor to study the ground
state phase diagram of a half filled Hubbard-Heisenberg model in a
square lattice with nearest neighbor hopping $t$ and a diagonal
hopping $t'$. The calculations are carried out by using
variational Monte Carlo method which treats the Gutzwiller
projection explicitly. At large on-site Coulomb interaction $U$, the
ground state is antiferromagnetic. As $U$ decreases, the ground
state becomes superconducting and eventually metallic. The phase
diagram is obtained by extensive calculations. As compared to the
strong effect of $U/t$, the phase boundaries turn out to be less
sensitive to $t'/t$. The result is consistent with the phase
diagram in layered organic conductors, and is compared to the
earlier mean field result based on the Gutzwiller approximation.

\end{abstract}

\maketitle

\section{INTRODUCTION}

High-temperature superconductivity remains to be an exciting and rich field. One
of the interesting proposals is Anderson's resonating valence bond (RVB)
state.\cite{AN1,AN2,Lee}
In the RVB theory, the parent compound is an insulator at half electron filling, or
one electron per Cu-site, and
chemical doping is essential to introduce charge carriers to lead to
superconductivity. The mathematics of the RVB theories therefore is
in a Hilbert space which completely projects out the on-site double-occupied electron states.
At the half filled, there is exactly one electron per lattice site,
and the charge degree of freedom is totally frozen, resulting in a
Mott-insulator.

Another interesting class of materials in the context of strongly
correlated systems is the layered organic
conductors,\cite{Jerome,MCKENZIE,ISHIGURO,LANG} which may undergo
a phase transition from an insulator to a superconducting (SC)
state by applying pressure.\cite{Jerome} Since these materials are
effectively at half filling,\cite{Fukuyama} the phase transition
is due to the competition between the Coulomb interaction and
kinetic band width, the latter of which is tuned by pressure
instead of chemical doping. There have been several related
theoretical works on layered organic superconductors in recent
years.\cite{Fukuyama,Fukuyama2,OT3,GAN1,OT2,OT1,OT4,TRI,MCKENZIE2,OT7,OT5,OT6} 
The mathematics of the
SC state may be described by a partially Gutzwiller projected BCS
state,\cite{FC1,GAN1} instead of the complete projection as in the
RVB theory. We shall refer to this partially Gutzwiller projected
BCS state as a Gossamer superconductor, a phrase first introduced
by Laughlin\cite{LAU1,LAU3} originally in the context of high
temperature superconductors. Gossamer superconductivity refers to
those SC states with a dilute superfluid-density. The partial
Gutzwiller projection allows charge fluctuations even at half
filling. One of us\cite{FC1} proposed that in this case an
effective model is the Hubbard-Heisenberg model which includes
the standard kinetic energy, the on-site Coulomb repulsion, as
well as the anti-ferromagnetic spin-exchange. The idea was applied
to the study of $\kappa$-(BEDT-TTF)$_2X$ by Gan {\it et al},\cite{GAN1}, where
the Gutzwiller approximation was used to replace the partial
Gutzwiller projection by a set of renormalized factors and the
resulted renormalized Hamiltonian was then studied by a mean field
theory. The finding is a phase-diagram distinguishing three
phases: normal metal, superconductor and anti-ferromagnet.

In this paper, we shall study the phase diagram of an effective
Hubbard-Heisenberg model by using variational Monte Carlo (VMC)
method. The order-parameters for the d-wave superconductivity and
for the anti-ferromagnetism are calculated directly. We obtain a
phase diagram consistent with the experiments, providing further
support to the scenario of the Gossamer superconductivity to
describe the layered organic conductors. Interestingly, our
numerical calculation of the SC order parameter suggests a
relatively high superfluid density near the phase boundary to the
AFM insulator. The results from our VMC
calculations also provide support to the earlier mean field
results based on the Gutzwiller approximation,\cite{GAN1} although
we find a less sensitive role of $t'$.

\section{MODEL, TRIAL-WAVE-FUNCTION AND METHOD}

\begin{figure}[tbp]
     \centering
    \includegraphics[width=0.6\columnwidth]{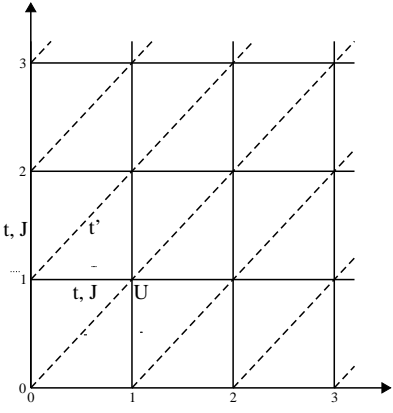}
    \caption{(Color online) Illustration of the lattice for Hamiltonian (1) studied in this paper.
t and J are hopping and spin exchange coupling between nearest neighbor pairs (solid lines),
and t' is the hopping integral along a diagonal direction (dashed lines).}
    \label{lattice}
\end{figure}

We study a Hubbard-Heisenberg model in a 2-dimensional lattice
illustrated in Fig. \ref{lattice}. The Hamiltonian is given by

\begin{eqnarray}
H=\sum_{i} Un_{i\uparrow} n_{i\downarrow} - \sum_{\langle i,j\rangle \sigma} t_{ij}
c_{i\sigma}^{\dagger} c_{j\sigma} + h.c. +
\sum_{(i,j)} J\vec{S_i} \cdot \vec{S_j}
\label{ham}
\end{eqnarray}
Here $c_{j\sigma}$ is the electron annihilation operator of an
electron with spin $\sigma$ on site $i$, $\vec S_i$ is the
spin-1/2 operator at site $i$, and
$n_{i\sigma}=c_{i\sigma}^{\dagger} c_{i\sigma}$. The sum $(i,j)$
is over the nearest neighbors (n.n.) pairs on the square lattice,
and the sum over $\langle i,j\rangle$  is over both the n.n. pairs
and the diagonal bonds (dashed lines in Fig. 1).  We set the n. n.
hopping $t_{i,j}=t=1$ as the energy unit, and fix the spin
exchange $J$ as $J/t=0.5$, and treat the diagonal hopping integral
$t_{i,j}=t'$ and the on-site repulsion $U$ as tuning parameters.
In our numerical calculations, we consider a $L$ by $L$ lattice
and use a periodic boundary condition along the x-direction and
antiperiodic boundary condition along the y-direction.

Our trial wave-function reads

\begin{equation}
\vert \Psi \rangle = e^{\beta \sum_{\langle i,j \rangle} S_i^z S_j^z }
\prod_i  (1-\alpha n_{i\uparrow} n_{i\downarrow}) \vert \Psi_{N_e} \rangle
\label{be}
\end{equation}
where $\vert \Psi_{N} \rangle$ is the BCS-wave function projected to the
subspace with the fixed number of particles $N_e$ as defined in Eq. (3) below.
In Eq. (2), we introduce two variational parameters $\alpha$ and $\beta$,
to control the partial Gutzwiller projection and the AFM correlation,
respectively. The BCS-wave function in the fixed particle formalism has the following form in real space:

\begin{eqnarray}
\vert \Psi_{N_e} \rangle  =  \left( \sum\limits_{j_\downarrow,l_\uparrow}
a(R_{j\downarrow}-R_{l\uparrow}) c_{l,\uparrow}^{\dagger} c_{j,\downarrow}^{\dagger}
\right)^{N_e/2}\vert 0 \rangle
\label{a1}
\end{eqnarray}
where $R_{j\sigma}$ is the spatial position of an electron with
spin $\sigma$ at the lattice site $j$, and the sum is over all the
pairs of a spin-up electron at the site $j$ and a spin-down
electron at site $l$. Here $a(\rb)$ is the amplitude of the wave
function, which is the Fourier transform of
$a(\kb)=v_{\kb}/u_{\kb}$, with $u_{\kb}$ and $v_{\kb}$ given in
the usual BCS wavefunction:

\begin{equation}
\vert \Psi_{BCS} \rangle = \prod_{\kb} (u_{\kb} + v_{\kb}  c_{\kb \uparrow}^{\dagger} c_{-\kb \downarrow}^{\dagger}) \vert 0 \rangle
\end{equation}
We have\cite{GROS1,AN1}:
\begin{eqnarray}
a(\rb) &= &\sum_{k} a_{\kb} \cos(\kb \rb) \label{a2} \nonumber \\
a_{\kb}: &= &\frac{\Delta(\kb)}{\xi_{\kb}+\sqrt{\xi_{\kb}^2+\Delta(\kb)^2}}
\end{eqnarray}
Following the previous literature on the pairing symmetry for the model
\cite{GAN1,OT3}, we focus here on the
$d_{x^2-y^2}$-wave pairing state, where $\Delta(\kb)$ and $\xi_{\kb}$ have the following forms,

\begin{equation}
\Delta(\kb)=\Delta (\cos(k_{x})-\cos(k_{y}))
\label{del}
\end{equation}
\begin{equation}
\xi_{\kb}=-2t(\cos(k_{x})+\cos(k_{y}))-2t'_v\cos(k_{x}+k_{y})-\mu
\label{dis}
\end{equation}
where $\Delta$, $\mu$ and $t'_v$ are variational parameters in the
theory. Note that $t'_v \neq t'$ in general due to the spin
coupling term in the Hamiltonian. The advantage of the above trial
wave function is that the SC and AFM order can be treated on an
equal footing. It turns out that a small value of $\beta$ improves
the energy of the SC state, while a sufficiently large value of
$\beta$ leads to AFM long range ordering. We measure the staggered
magnetization to quantitatively study the the AFM phase,

\begin{equation}
m = \sqrt{\frac{1}{N(N-1)} \sum_{\langle i,\vec{r}\rangle} \langle S_i^z S_{i+\vec{r}}^z \rangle (-1)^{r_x+r_y}}
\label{afop}
\end{equation}
where $N$ is the number of the lattice sites, and the sum is over all the $N(N-1)$ pairs between sites $i$ and $i+\vec r$
on the lattice.
To measure the SC long range order, we introduce a pair correlation function:

\begin{eqnarray}
\phi_{i,j} &=&   F_i F^{\dagger}_j \nonumber\\
F_i &=& \frac{1}{4} \sum_{\tau} b_{i,i+\tau} (-1)^{\tau_y} \nonumber\\
b_{i,i+\tau} &=& \frac{1}{\sqrt 2}(c_{i\downarrow} c_{i+\tau \uparrow}- c_{i\uparrow} c_{i+\tau \downarrow}),
\label{odlro}
\end{eqnarray}
where $b_{i,i+\tau}$ is a spin singlet bond between the two sites $i$ and $i+\tau$, and $\tau= \pm \hat x, \pm \hat y$,
$\tau_y =0$ for $\tau=\pm \hat x$, and $\tau_y =\pm 1$ for $\tau=\pm \hat y$.
$F_i$ describes a d-wave singlet bond around the site $i$.
The off-diagonal long range order parameter for the d-wave pairing can be measured by the quantity at $R \rightarrow
\infty$,
\begin{eqnarray}
\phi(\vec R) = \frac{1}{N}\sum_{i}\langle \phi_{i, i+ \vec R} \rangle,
\end{eqnarray}
where the sum is over all the lattice sites. In our calculations on the finite size systems, we choose
$\vec R =(L/2, L/2)$, the largest displacement on the lattice of $L$ by $L$ with $L$ upto 10.

To simplify the variational procedure, we will fix $\mu$ in the
calculations with the reasons given below. It has been argued
\cite{GROS1} that $\Delta$ and $\mu$ are not independent in the
variational calculations for the $t-J$ model. We found that the
results are essentially insensitive to $\mu$ for the present
model. In contrast, $t'_v$ is an important variational parameter
here. The ground state energies and the ground state phase is
sensitive to $t'_v$ over a wide parameter-range. By fixing $\mu$,
we have then four variational parameters ($\Delta, t'_v, \alpha,
\beta$) in our calculations to determine the phase diagram in the parameter space of $U$ and $t'$.\\

There are two sources of error bars in our numerical calculations
within the variational approach. One is from the statistical
errors, and the other is due to the discreteness of the
variational parameters in our calculations. In our simulation, we
start with several different initial configurations and then
average our numerical measurements over those simulations. The
error-bars obtained in these averages are found to be one order of
magnitude smaller than the error-bars described below. We consider
the possible values for the variational parameters and divide them
into small slices. Then we perform VMC for all combinations of
this "mesh". After obtaining an optimal set of variational
parameters in this mesh for a particular set of $U/t$ and $t'/t$,
we develop a local mesh for nearby values of the tuning
parameters. From the spacing of our mesh, we obtain the error-bars
for the variational parameters. The results and the error-bars we
present in this paper are essentially due to the finite elements
we choose in the variational parameters.

\section{RESULTS AND DISCUSSION}

\begin{figure}[tbp]
     \centering
    \includegraphics[width=0.9\columnwidth]{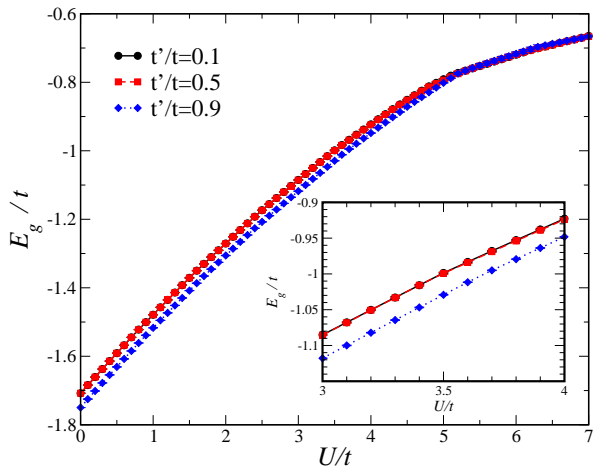}
    \caption{(Color online) Variational ground state energy $E_g$ of Hamiltonian (1) as a function of $U/t$ for
$J/t=0.5$ and various values of $t'/t$. The inset is an enlarged
figure for the energy.}
    \label{GLOBAL}
\end{figure}

\begin{figure*}[ht!!!]
     \centering
     \vspace{-.5cm}
    \includegraphics*[width=16cm]{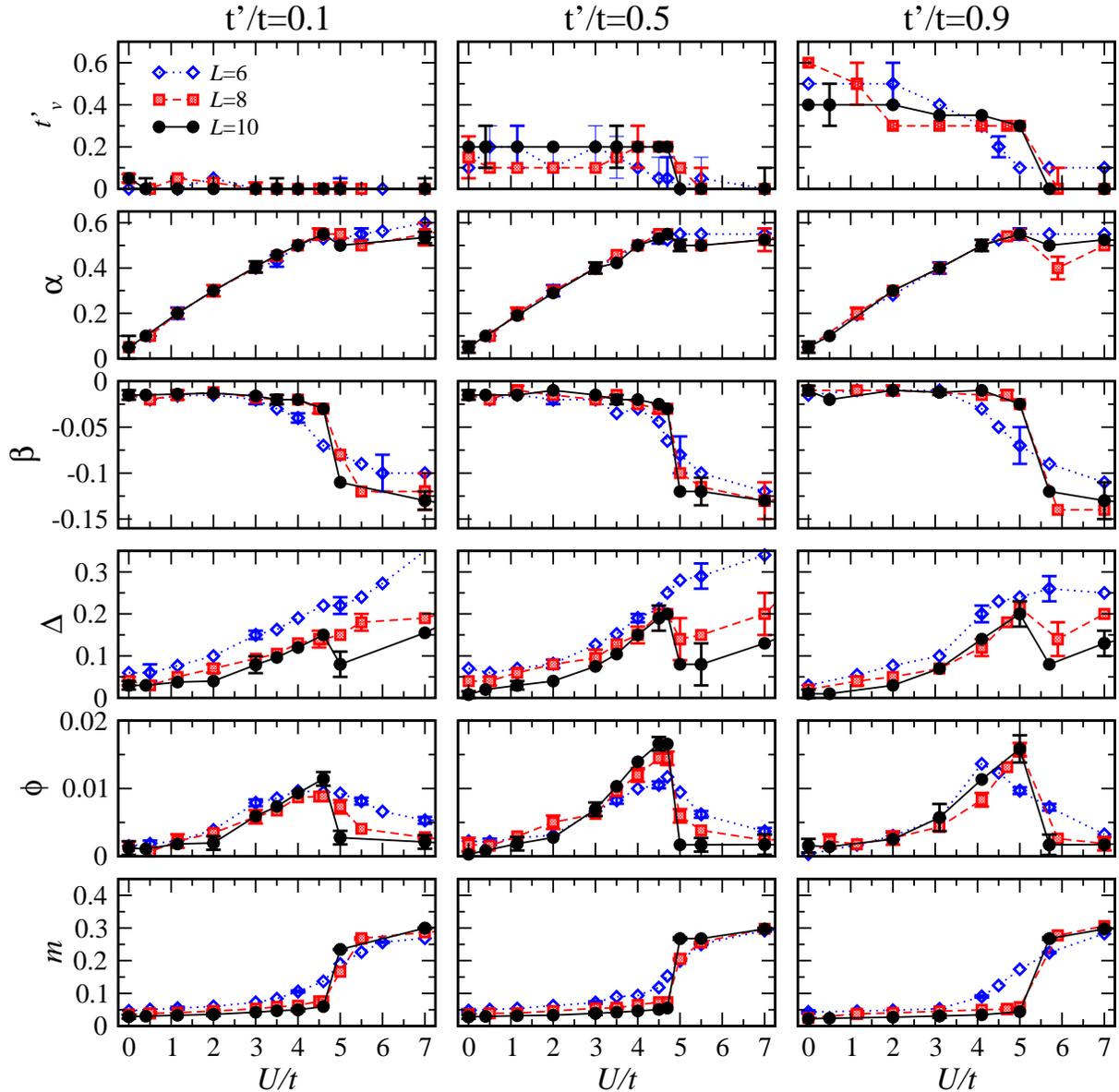}
   \caption{(Color online) The ground state variational parameters $t'_v$ (Eq. \ref{dis}),
$\alpha$ (partial projection, Eq. \ref{be}), $\beta$ (AFM weighting, Eq. \ref{be}) and
$\Delta$ (pairing amplitude, Eq. \ref{del}) as functions of $U/t$ for $t'/t=0.1$ (left),
$t'/t=0.5$ (mid) and $t'/t=0.9$ (right).
$J/t =0.5$ is fixed.
Also plotted are the measured d-wave SC order parameter $\phi$ (Eq. \ref{odlro})
and the staggered magnetization $m$ (Eq. \ref{afop}). The lattice size is $L \times L$, with
$L=6,\, 8, \, 10$.   The selected error bars shown are typical, due to the
finite parameter spacing in our calculations.}
    \label{bigpicture}
\end{figure*}

\begin{figure}[tbp]
    \subfigure[ ]{
    \label{fig:mini:subfig:a}
    \begin{minipage}[t]{.24\textwidth}
    \centering
    \includegraphics[width=\textwidth]{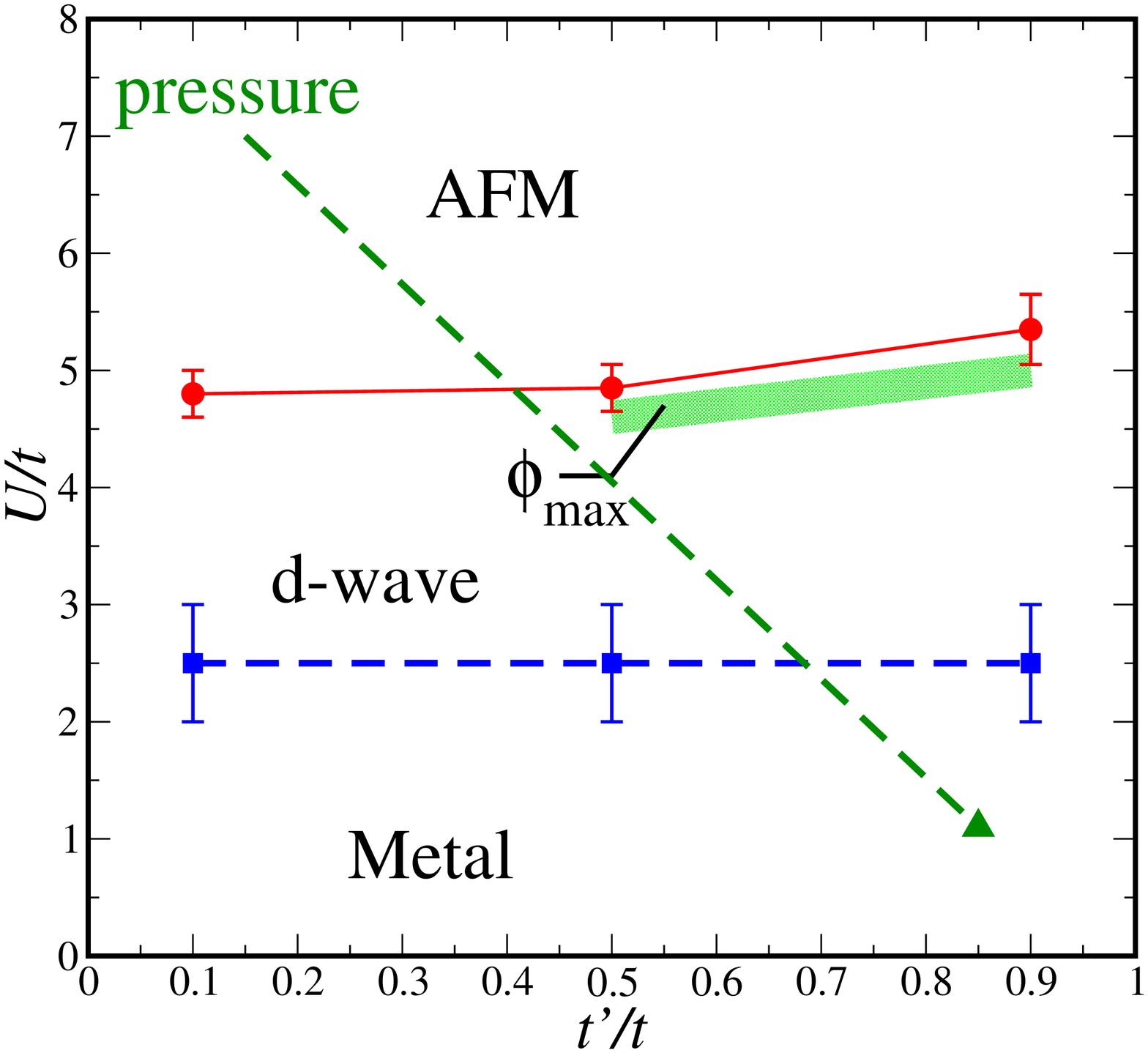}
    \end{minipage}}%
    \subfigure[ ]{
    \label{fig:mini:subfig:b}
    \begin{minipage}[t]{.24\textwidth}
    \centering
    \includegraphics[width=\textwidth]{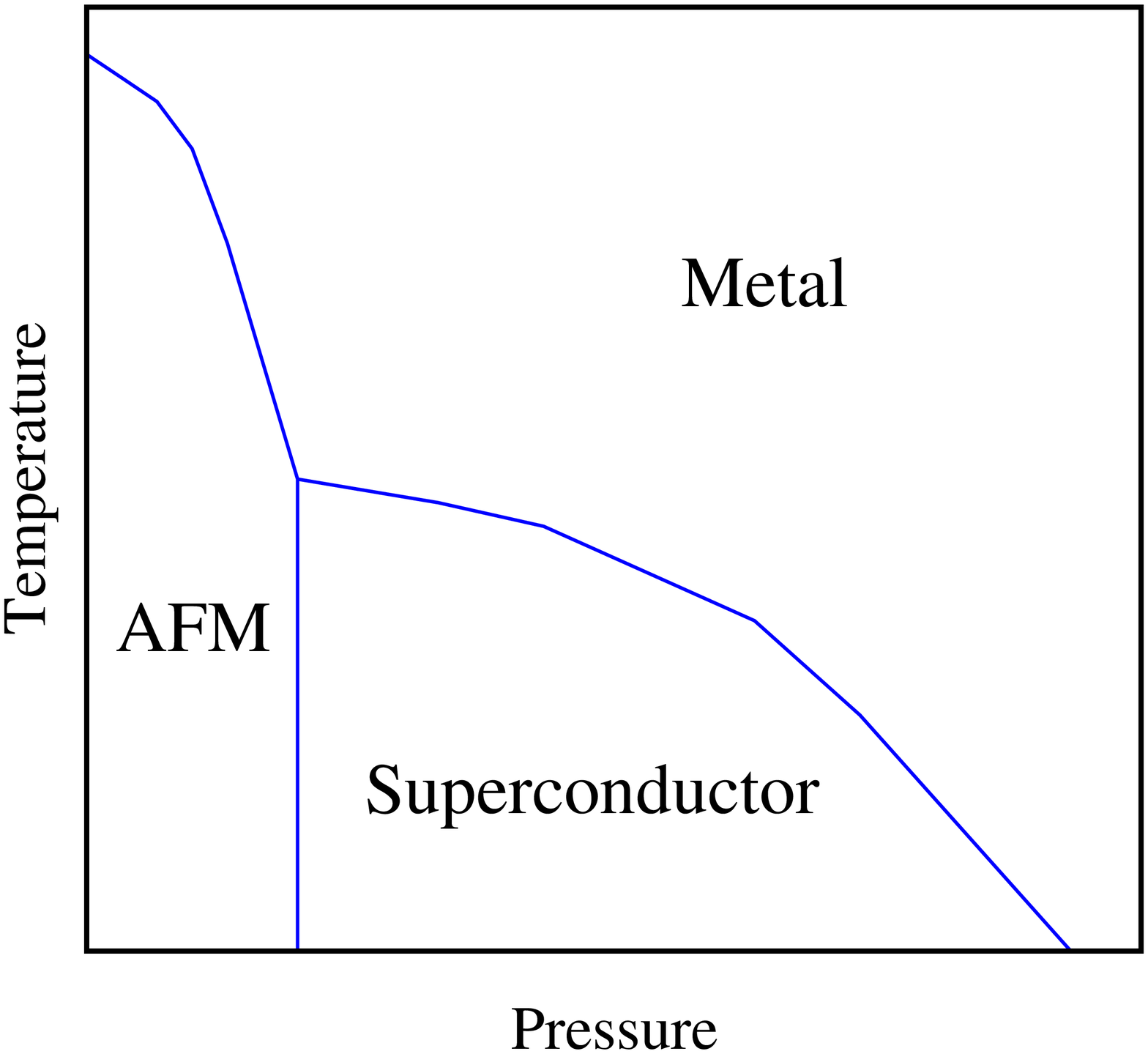}
    \end{minipage}}
    \caption{(Color online) (a) Phase diagram of the ground state of the Hamiltonian (1)
obtained from our VMC calculations: Metal (metallic phase), d-wave
SC (superconducting phase), and AFM (anti-ferromagnetic phase).
The region with large SC order parameter is indicated by a thick
line and marked with $\phi_{max}$. The values of the
order-parameters for the d-wave and for the AFM phases can be
found in Fig. \ref{bigpicture}. The arrow indicates the schematic
flow of the parameters when pressure is applied. (b) Schematic
phase diagram of organic superconductors in parameter space of
temperature and pressure.} \label{gossphase}
\end{figure}
In this section, we present our results on the variational ground state, the corresponding variational
parameters, the SC and AFM long range orders of the Hamiltonian (1) in the parameter space of $U/t$ and
$t'/t$.  Since the phase of the ground state is much more sensitive to the on-site repulsion
$U$ than to the diagonal hopping $t'$, we focus our study on three values of $t'$, with $t'/t=0.1,0.5,0.9$.
In Fig. \ref{GLOBAL} we plot the obtained ground state energies
as functions of $U$ for different values of $t'$.  The corresponding optimized variational parameters
and long range SC and AFM order parameters as functions of
$U$ for three sets of values of $t'$ are plotted in Fig. \ref{bigpicture}.
The simulations are carried out on lattice sizes of L=6, 8, and 10,
as indicated in the figure.  Before we discuss the results, we note that the spin
coupling term in Eq. (1) is to account the virtual hopping process
in the Hubbard model, which is derived at the large $U$ limit. The present study may be of relevance to the
Hubbard model only at large $U$, but not at small $U$.  Our main interest will be
at large or intermediate values of $U$, and the interpretation of our
results at small $U$ to the Hubbard model should be cautious.

Before we discuss general features, we briefly  discuss the
obtained variational parameter $t'_v$, which is to optimize the
kinetic energy due to the presence of the diagonal hopping
integral $t'$. $t'_v$ increases as $t'$ increases, but $t'_v$ is
significantly smaller than $t'$ as we can see from the first row
in Fig. \ref{bigpicture}. At large $U$, $t'_v$ becomes zero or very tiny. This may
be understood as a result of the AFM ground state with
commensurate wave vector $(\pi,\pi)$, since a finite $t'_v$ does
not match the AFM state and is not preferred.

As $U$ increases from zero, the projection parameter $\alpha$
increases from around $0.05$, indicating a graduate increase in
Gutzwiller projection, while the weighting  factor parameter
$\beta$ changes little at small $U$, but changes rapidly around $U\approx 5$. 
The mean field pairing amplitude parameter $\Delta$
changes slowly at small $U$, but increases rapidly starting from around
$U=2$, then reaches a maximum at around $U=4.5$ and drops at
larger $U$. The ground state properties are best seen in the
measurement of the SC order parameter $\phi$ and AFM order
parameter $m$. Qualitatively there are three regions as $U$
increases. At small $U$, both $\phi$ and $m$ are very tiny or
essentially zero, indicating a metallic state. At intermediate
$U$, $\phi$ increases monotonically with $U$, while $m$ remains
tiny, indicating a SC state. As $U$ further increases, $\phi$
drops sharply, while $m$ increases rapidly. We may identify this
phase as the AFM phase without the SC order. The tiny but non-zero
values of $\phi$ and $m$ in the non-ordered states may be
explained as finite size effect, although a systematic scaling
analyses is difficult due to the small sizes we have studied. The
above features are qualitatively similar for $t'/t=0.1, \, 0.5,\,
0.9$. This is somewhat different from the early analytic
calculations by using Gutzwiller approximations on the projected
wavefunctions, where $t'$ is found to suppress the AFM phase. We
note that while the onset for the SC-phase is similar for
different $t'$, the magnitude of the SC-order parameter is much
bigger for the $t'/t=0.5$ and $t'/t=0.9$. As we can see from the
figure, the largest SC order parameter $\phi$ is found near the
boundary to the AFM-phase. At $t'=0$, we expect the model (1) to
have instability towards a commensurate AFM state for any finite
$U$.

In Fig. \ref{gossphase}, we plot the phase diagram of model (1)
obtained within our variational wavefunctions. While the
phase-boundary between SC and AFM can be found easily by
considering one point clearly belonging to the AFM and one point
clearly belonging to the SC phase, between SC and metallic phase we
have to use an arbitrary value to define the phase-boundary, as
the onset of the SC-order parameter $\phi$ is not so sharp. We
choose $\phi < 0.004$ as our criteria classifying the phase to
be SC. The error-bars in this diagram reflect within which area we
have uncertainty that a point would be in either of the two phases
considered. Comparing the phase diagram obtained in the VMC method
with the previous result by using renormalized mean field
theory,\cite{GAN1} they  qualitatively agree with each other in
the sense both give the three phases, and overall features are
similar. However, there are two differences.  First, while both of
the methods give the transition point between the SC- and
AFM-phases to increase when $t'/t$ increases, in the VMC
calculation, the effect is not as big as in the earlier
Gutzwiller-approximation based calculation. In our calculation if
we consider a fixed and non-zero $t'_v$ instead of a variational
one, we would in fact get a slope close to the one reported by Gan
{\it et al}. The second difference is that our VMC suggests the
onset for superconductivity to be at $U=2.5$ for all cases, while
Gan {\it et al} find this phase boundary changing considerable
when tuning $t'/t$. We believe that these differences can be
attributed to the different method and wave-function used. For
comparison with the experiments, we plot a schematic phase diagram
for the layered organic conductors at the right panel of the
figure.

In summary, we have presented the results of VMC calculations for
a recently suggested model for Gossamer superconductivity. Our
trial wave function has the ingredient to describe metallic, AFM
and SC states. This was archived by means of using Jastrow-factors
for partial Gutzwiller-projection and AFM-weighting. We showed
that the VMC result is consistent with experiments, and supports
the previously suggested analytical variational calculations
qualitatively, as we were able to identify the three expected
phases, with the help of measurements of the order-parameters for
AFM and SC. The exact transition line between SC- and
metallic-phase, and between SC- and AFM-phase differs from the one
found previously.

\acknowledgements

The VMC calculations have been carried out on the
HPCPOWER-cluster, the WinHPC-cluster and the windows-condor all
powered by HKU's computer centre. We wish to thank Masao Ogata for
advise on VMC. Further we wish to thank Kwan Wing Keung for help
in parallelization of the code, and customization of the code for
the windows-condor-system. The work was partly supported by Hong
Kong's RGC grant. The work in Nanjing was supported by NSFC
10325416, the Ministry of Science and Technology of China (under
the Grant No. 2006CB921802 and 2006CB601002) and the 111 Project
(under the Grant No. B07026).

\end{document}